\documentclass[aps,prl,letterpaper,twocolumn,showpacs,superscriptaddress,floatfix,longbibliography]{revtex4-1}
\usepackage[english]{babel} \usepackage{amsmath} \usepackage{color}
\usepackage{epsfig} \usepackage{graphicx} \usepackage{bm}
\usepackage{mathtools}

\usepackage{times,amsmath,amssymb} \usepackage{amsfonts}
\usepackage{amssymb}

\usepackage[colorlinks,urlcolor=blue,bookmarks=false,hypertexnames=true]{hyperref}

\allowdisplaybreaks 
\newcommand{\E}{\mbox{$\mathsf E$}}
\newcommand{\media}[1]{\left\langle #1 \right\rangle}
 \newcommand{\ket}[1]{| #1 \rangle}
\newcommand{\scp}[2]{\langle #1 | #2 \rangle}
\newcommand{\braket}[3]{\langle #1 | #2 | #3 \rangle}

\newcommand{\Span}{\mathrm{span}} %

\DeclareMathOperator{\TDlim}{TD-lim}

\newcommand{\eps}{\epsilon} 
\definecolor{myred}{RGB}{168,5,14}
\definecolor{myblue}{RGB}{13,13,255}
\definecolor{editorcolor}{RGB}{168,5,14}
\definecolor{mygreen}{RGB}{20,150,20}

\begin{document}

\title{Wigner Crystallization of Electrons in a One-Dimensional
  Lattice:\\ A Condensation in the Space of States}

\author{Massimo Ostilli} \affiliation{Instituto de F\'isica,
  Universidade Federal da Bahia, Salvador 40170-115, Brazil}

\author{Carlo Presilla} \affiliation{Dipartimento di Fisica, Sapienza
  Universit\`a di Roma, Piazzale A. Moro 2, Roma 00185, Italy}
\affiliation{Istituto Nazionale di Fisica Nucleare, Sezione di Roma 1,
  Roma 00185, Italy}

\date{\today}

 \begin{abstract}
   We study the ground state of a system of spinless electrons
   interacting through a screened Coulomb potential in a lattice ring.
   By using analytical arguments, we show that, when the effective
   interaction compares with the kinetic energy, the system forms a
   Wigner crystal undergoing a first-order quantum phase transition.
   This transition is a condensation in the space of the states and
   belongs to the class of quantum phase transitions discussed in
   J. Phys.~A \textbf{54}, 055005 (2021).  The transition takes place
   at a critical value ${r_s}_{c}$ of the usual dimensionless
   parameter $r_s$ (radius of the volume available to each electron
   divided by effective Bohr radius) for which we are able to provide
   rigorous lower and upper bounds. For large screening length these
   bounds can be expressed in a closed analytical form.  Demanding
   Monte Carlo simulations allow to estimate
   ${r_s}_{c}\simeq 2.3 \pm 0.2$ at lattice filling $3/10$ and
   screening length 10 lattice constants. This value is well within
   the rigorous bounds $0.7\leq {r_s}_{c}\leq 4.3$.  Finally,
   we show that if screening is removed after the thermodynamic limit
   has been taken, ${r_s}_{c}$ tends to zero. In contrast, in a
   bare unscreened Coulomb potential, Wigner crystallization always
   takes place as a smooth crossover, not as a quantum phase
   transition.
 \end{abstract}

 \maketitle
 The Wigner crystal (WC)~\cite{Wigner}, namely, the periodic
 arrangement of electrons that minimizes the Coulomb interaction
 energy in the presence of band motion effects~\cite{Hubbard1978}, has
 been investigated in several long-range repulsive potential
 models~\cite{Sinai1983,Bak1982,Fratini2004,Slavin2005}. Two
 dimensional~\cite{2DEG_1989,2DEG_1996,
   2DEG_2002,2DEG_2009,2DEG_2017,Noda2002} and
 one-dimensional~\cite{Siegmund2009,VFB2003} electron gases at zero
 temperature have been extensively studied from a theoretical point of
 view.  A recent experiment succeeded in imaging an electronic WC in
 one-dimensional nanotubes~\cite{nanotube}.

 The occurrence of a WC is often argued by comparing the typical
 kinetic and Coulomb energies involved. Roughly speaking, the kinetic
 energy can be evaluated as $\hbar^2/(2 m^* r^2)$, where $m^*$ is the
 effective electron mass and $r$ the radius of the volume available to
 each electron, whereas the Coulomb energy can be taken as $e^2/r$,
 where $e$ is the electron charge. These two energies have the same
 value when $r_s \equiv r/a_{B}$, $a_{B}$ being the
 effective Bohr radius, is equal to the ``critical value''
 ${r_s}_{c}=2$.  Then one concludes that for
 $r_s > {r_s}_{c}$ a WC must show up.

 The above argument can be, however, misleading.  Consider the case of
 the unscreened Coulomb potential in a $d$-dimensional space with a
 fixed value of $r_s$.  For a gas of $N_{p}$ electrons, the
 energy per particle of the \emph{bare} $d$-dimensional Coulomb
 potential scales as ${N_{p}}^{d-1}$ for $d>1$, and as
 $\ln N_{p}$ for $d=1$~\cite{Dubin}. On the other hand, at any
 dimension $d$, the kinetic energy per particle is independent of
 $N_{p}$, so that the potential energy overwhelms the kinetic
 one for $N_{p}$ large enough.  In other words: in the
 thermodynamic limit ($\TDlim$), ${r_s}_{c}\to 0^+$ and no
 quantum phase transition (QPT) takes place, the system being
 trivially a WC for any $r_s> 0$; for finite $N_{p}$, instead,
 the transition from free electron motion to WC obtained by increasing
 $r_s$ is just a smooth crossover, not a QPT.

 Screening is, therefore, an essential ingredient~\cite{Hubbard1978}:
 the ground-state (GS) energy per particle of the screened potential
 scales linearly with $N_{p}$ and can fairly compete with the
 kinetic term.  It is only in this case that we can hope to observe a
 QPT in the $\TDlim$ by varying $r_s$.

 We are not aware of any conclusive study on the phase transition
 nature of the Wigner crystallization, except for the work of Brascamp
 and Lieb on the $1d$ plasma in a neutralizing
 background~\cite{Lieb2002}.  Here, we study the ground state of a
 system of spinless electrons interacting through a screened $3d$
 Coulomb potential in a lattice ring. By using analytical arguments,
 we demonstrate that, for any finite screening length, the Wigner
 crystallization is a QPT taking place at a finite critical value
 ${r_s}_{c}$ of the parameter $r_s$.  For ${r_s}_{c}$ we
 provide rigorous upper and lower bounds, which can be cast in an
 analytical form in the limit of large screening length.  The QPT that
 we find is of first order (according to Ehrenfest classification) and
 falls within the class of condensations in the space of states
 introduced in~\cite{QPT}.  Demanding Monte Carlo (MC) simulations
 based on an advanced bias-free code~\cite{MC} allow to estimate a
 value of ${r_s}_{c}$, which is well within the rigorous
 bounds.  Finally, we show that, removing the screening after the
 $\TDlim$ has been taken, we have ${r_s}_{c}\to 0^+$,
 confirming that a nonzero minimal screening is necessary to have a
 realistic physical picture.

 We briefly recall the mechanism of first-order QPT of~\cite{QPT}.  To
 be specific, let us consider a lattice model with $N$ sites and
 $N_{p}$ particles described by a Hamiltonian
 \begin{align}
   \label{H}
   H=K+g V,
 \end{align}
 where $K$ and $V$ are Hermitian noncommuting operators, and $g$ a
 free dimensionless parameter, which, without loss of generality, can
 be taken to be non-negative.  Regardless of the details of $K$ and
 $V$, we represent $H$ in the eigenbasis of $V$ and it is natural to
 call $V$ the potential operator, and $K$ the hopping operator. To
 exclude trivial behaviors, we suppose that the eigenvalues of $K$ and
 $V$ scale linearly with the number of particles $N_{p}$.
 Since in the two opposite limits $g\to 0$ and $g\to \infty$, the GS
 of the system tends to the GS of $K$ and $V$, respectively, we wonder
 if, in the $\TDlim$, this transition occurs as a QPT taking place at
 some critical value $g_{c}$.

 A quite general kind of QPT is the condensation in the space of
 states.  We decompose the Hilbert space $\mathbb{F}$ of the system as
 the direct sum of two mutually orthogonal subspaces, denoted
 condensed and normal, namely,
 $\mathbb{F}=\mathbb{F}_\mathrm{cond} \oplus
 \mathbb{F}_\mathrm{norm}$.  The definition of these subspaces is as
 follows.  We write $\mathbb{F} = \Span \{ \ket{n} \}_{n=1}^{M}$,
 where $\{ \ket{n} \}$ (later on called configurations) is a complete
 orthonormal set of eigenstates of $V$, i.e., we have
 $V \ket{n} =V_n \ket{n}$, $n=1,\dots,M$, where we assume ordered,
 possibly degenerate, potential values
 $V_1 \leq V_2 \leq \dots \leq V_M$. Given an integer
 $M_\mathrm{cond}<M$, we then define
 $\mathbb{F}_\mathrm{cond} = \Span \{ \ket{n}
 \}_{n=1}^{M_\mathrm{cond}}$ and
 $\mathbb{F}_\mathrm{norm} = \Span \{ \ket{n}
 \}_{n=M_\mathrm{cond}+1}^{M} = \mathbb{F}_\mathrm{cond}^\perp$.  This
 definition essentially relies on the choice of the dimension
 $M_\mathrm{cond}$, which, in view of the ordering of the potential
 values, marks the maximum potential value included in the condensed
 subspace
 \begin{align}
   \label{maxVcond}
   \max V_\mathrm{cond} =
   \max\{V_n:~\ket{n}\in\mathbb{F}_\mathrm{cond}\} =
   V_{M_\mathrm{cond}}.
 \end{align}
 Consider the GS energies of the system, the condensed, and normal
 subspaces:
 \begin{align}
   E &=\inf_{\ket{u}\in\mathbb{F}} \braket{u}{H}{u}/\scp{u}{u},
   \\
   \label{Econd}
   E_\mathrm{cond} &= \inf_{\ket{u}\in\mathbb{F}_\mathrm{cond}}
                     \braket{u}{H}{u}/\scp{u}{u},
   \\
   E_\mathrm{norm} &= \inf_{\ket{u}\in\mathbb{F}_\mathrm{norm}}
                     \braket{u}{H}{u}/\scp{u}{u}.
 \end{align}
 We are interested in the situations where $M_\mathrm{cond}/M\ll 1$
 and, as a consequence,
 $M_\mathrm{norm}/M\equiv (M-M_\mathrm{cond})/M\simeq 1$.
 % By contrast, $M_\mathrm{norm}=(M-M_\mathrm{cond})/M\simeq 1$.
 This justifies the names \textit{condensed} and \textit{normal}
 assigned to the two subspaces and suggests the following dichotomy
 argument: since $\mathbb{F}\simeq\mathbb{F}_\mathrm{norm}$, we have
 $E\simeq E_\mathrm{norm}$---unless---it is energetically more
 convenient to ``freeze'' into the infinitely smaller subspace
 $\mathbb{F}_\mathrm{cond}$, where we get $E\simeq E_\mathrm{cond}$.

 The above heuristic argument can be cast in rigorous terms as
 follows.  The $\TDlim$ is defined as the limit
 $N,N_{p}\to\infty$ with $N_{p}/N=\varrho$ constant.
 Consider the rescaled energies:
 \begin{align}
   \epsilon(g) &= \TDlim E(N,N_{p},g)/N_{p},
   \\
   \epsilon_\mathrm{cond}(g)
               &=\TDlim E_\mathrm{cond}(N,N_{p},g)/N_{p},
   \\
   \epsilon_\mathrm{norm}(g)
               &=\TDlim E_\mathrm{norm}(N,N_{p},g)/N_{p},
 \end{align}
 which are finite in view of the assumed scaling properties of $K$ and
 $V$ (dependence on $\varrho$ is left understood).  In~\cite{QPT} we
 have proved the following general theorem
 \begin{subequations}
   \begin{align}
     \label{QPT0}
     &\mbox{if}~ \TDlim {M_\mathrm{cond}/M}=0,
     \\
     \label{QPT1}
     &\mbox{then}~
       \epsilon = \min\{\epsilon_\mathrm{cond},\epsilon_\mathrm{norm}\}.
   \end{align}
 \end{subequations}
 This theorem establishes the possibility of a QPT between a normal
 phase characterized by the energy per particle
 $\epsilon_\mathrm{norm}$, obtained by removing from $\mathbb{F}$ the
 infinitely smaller subspace $\mathbb{F}_\mathrm{cond}$, and a
 condensed phase characterized by the energy per particle
 $\epsilon_\mathrm{cond}$, obtained by restricting the action of $H$
 onto $\mathbb{F}_\mathrm{cond}$.  The situation is particularly
 simple for systems characterized by a single parameter as in the case
 of Eq.~(\ref{H}).  If Eq.~(\ref{QPT0}) holds and, moreover, the
 functions $\epsilon_\mathrm{norm}(g)$ and $\epsilon_\mathrm{cond}(g)$
 are such that the equation
 \begin{align}
   \label{QCP}
   \epsilon_\mathrm{norm}(g)=\epsilon_\mathrm{cond}(g)
 \end{align}
 admits a unique \textit{finite} solution $g=g_{c}$,
 Eq.~(\ref{QPT1}) provides
 \begin{align}
   \label{QPT2}
   \epsilon(g) = \left\{
   \begin{array}{ll}
     \epsilon_\mathrm{norm}(g), \qquad&g<g_{c},
     \\
     \epsilon_\mathrm{cond}(g), \qquad&g>g_{c}.
   \end{array}
                                        \right.
 \end{align}
 Equations~(\ref{QCP}) and (\ref{QPT2}) imply the existence of a
 first-order QPT at the critical point $g_{c}$.  In fact,
 although in general $\epsilon_\mathrm{cond}(g)$ and
 $\epsilon_\mathrm{norm}(g)$ are separately analytic in
 $g=g_{c}$, on observing that $\epsilon_\mathrm{cond}(g)$ and
 $\epsilon_\mathrm{norm}(g)$ are different functions, we conclude
 that, while $\epsilon(g)$ is continuous at $g=g_{c}$, its
 first derivative undergoes the discontinuity
 $|\epsilon_\mathrm{cond}'(g_{c})-
 \epsilon_\mathrm{norm}'(g_{c})|>0$.

 Whereas Eq.~(\ref{QPT0}) can be checked easily, the existence of a
 finite solution to Eq.~(\ref{QCP}) can be difficult to prove.  A
 practical approach can be as follows.  For $N,N_{p}$ finite
 with $N_{p}/N=\varrho$ constant, we evaluate
 $g_\mathrm{cross}(N,N_{p})$ as the value of the parameter $g$,
 if any, solution of the equation
 \begin{align}
   \label{QCPfinite}
   E_\mathrm{norm}(N,N_{p},g)=E_\mathrm{cond}(N,N_{p},g).
 \end{align}
 Assuming a smooth limiting behavior, we expect
 \begin{align}
   \label{gc_gcross}
   g_{c} = \TDlim g_\mathrm{cross}(N,N_{p}).
 \end{align}
 Even if this limit cannot be exactly evaluated, as in the case of
 numerical simulations, Eq.~(\ref{gc_gcross}) can be used to provide
 strict upper and lower bounds to $g_{c}$ as shown ahead.

 To recapitulate, if we find a partition
 $\mathbb{F}=\mathbb{F}_\mathrm{cond} \oplus \mathbb{F}_\mathrm{norm}$
 such that Eq.~(\ref{QPT0}) and Eq.~(\ref{QCP}) are satisfied, then a
 first-order QPT of the type introduced in~\cite{QPT} occurs at
 $g=g_{c}$.  In general, such a partition is not unique. In
 fact, for Eq.~(\ref{QCP}) to admit a solution with condition
 (\ref{QPT0}) satisfied, $\mathbb{F}_\mathrm{cond}$ can invariantly be
 chosen provided that it is not too small and not too large in such a
 way that neither of the two restrictions of $H$, to
 $\mathbb{F}_\mathrm{cond}$ and to $\mathbb{F}_\mathrm{norm}$, have a
 QPT. In this case, $\epsilon_\mathrm{cond}$ and
 $\epsilon_\mathrm{norm}$ are both analytic functions of $g$ at
 $g=g_{c}$, whereas $\epsilon$ is not.  Note that, for finite
 sizes, different partitions of $\mathbb{F}$ lead, in general, to
 different values of both $E_\mathrm{cond}(g)$ and
 $E_\mathrm{norm}(g)$.  Only in the $\TDlim$ different invariant
 partitions of $\mathbb{F}$ lead to the same values of
 $\epsilon_\mathrm{cond}(g)$ for $g>g_{c}$ and
 $\epsilon_\mathrm{norm}(g)$ for $g<g_{c}$, namely,
 $\epsilon(g)$, as indicated by Eq.~(\ref{QPT2}).  We will exploit
 this invariance to get rigorous bounds to $g_{c}$.

 We apply the above general strategy to a system of $N_{p}$
 electrons interacting in a ring of $N$ sites. As usual, for
 simplicity and saving computational efforts, we consider spinless
 particles.  The electronic Hamiltonian $H_e$ cast in the
 dimensionless form (\ref{H}) by $H_e/t=H=K+gV$,
 $t=\hbar^2/(2 m^* a^2)$ being the hopping coefficient with $m^*$ the
 effective electron mass and $a$ the lattice constant~\cite{VFB2003},
 is given by
 \begin{align}
   \label{K}
   K &= -\sum_{i=1}^{N}
       \left(c^\dag_{i}c_{i+1}+c^\dag_{i+1}c_{i}\right),
   \\
   \label{V}
   V &= \sum_{i=1}^{N}\sum_{j=i+1}^{N}v_{i,j}
       c^\dag_i c_i c^\dag_j c_j,
 \end{align}
 where the fermionic annihilation operators obey the periodic
 condition $c_{i+N}=c_i$. We consider a screened Coulomb
 interaction~\cite{Hubbard1978}
 \begin{align}
   \label{vij}
   v_{i,j} = \frac{1}{d_{i,j}} e^{-ad_{i,j}/R},
 \end{align}
 $R$ being the screening length and
 $d_{i,j}=\min(j-i,N+i-j), \quad j>i$, the dimensionless distance
 between sites $i$ and $j$ in the ring.  Screening takes into account
 the many-body effects not explicitly considered in $H$ and allows for
 the interaction energy to scale linearly with the number of particles
 $N_{p}$, as physically expected.  The value of $R$ depends on
 the microscopic details of the system considered.  However, whereas
 the minimum of $V$ has a logarithmic dependence on $R$, see later,
 the associated GS has a universal structure~\cite{Hubbard1978} under
 conditions on $v_{i,j}$~\cite{Sinai1983,Bak1982,Slavin2005} that are
 fulfilled by Eq.~(\ref{vij}) for any $R$.  With the above choice for
 the potential, the dimensionless coupling $g$ in Eq.~(\ref{H}) takes
 the form of the following Seitz radius~\cite{rs}
 \begin{align}
   \label{g}
   g = 2a/a_{B}, \qquad a_{B}= \hbar^2/(m^* e^2).
 \end{align}
 Now we determine a partition
 $\mathbb{F}=\mathbb{F}_\mathrm{cond} \oplus \mathbb{F}_\mathrm{norm}$
 which satisfies the conditions (\ref{QPT0}) and (\ref{QCP}).  We
 recall that, according to Eq.~(\ref{maxVcond}), a partition is
 defined by specifying the maximum potential value allowed in
 $\mathbb{F}_\mathrm{cond}$.

 As we show in~\cite{SM}, in the $\TDlim$ the distribution of the
 potential values (\ref{V}) divided by $N_{p}$ tends to a Dirac
 delta centered at $\overline{V}/N_{p}$, namely, the mean
 classical value of the potential per particle.  This implies that,
 whenever
 $\max V_\mathrm{cond}/N_{p} < \overline{V}/N_{p}$, we
 have $M_\mathrm{cond}/M \to 0$ in the $\TDlim$, i.e.,
 Eq.~(\ref{QPT0}) is satisfied.

 To comply with Eq.~(\ref{QCP}), consider that $E$, $E_\mathrm{cond}$,
 and $E_\mathrm{norm}$, are monotonously increasing functions of $g$
 convex upward~\cite{SM} and suppose that the critical point is
 unique.  It follows that $g_{c}$ is finite if and only if (i)
 $\epsilon_\mathrm{norm}(0) < \epsilon_\mathrm{cond}(0)$ and (ii)
 $\lim_{g\to\infty} \epsilon_\mathrm{cond}(g)/g < \lim_{g\to\infty}
 \epsilon_\mathrm{norm}(g)/g$.

 Condition (i) is equivalent to saying that in the $\TDlim$
 $\min K_\mathrm{norm}/N_{p} < \min
 K_\mathrm{cond}/N_{p}$.  Here and in the following, we use a
 notation as in Eq.~(\ref{maxVcond}), for example,
 $\min K_\mathrm{cond}$ is the smallest eigenvalue of the operator $K$
 restricted to the condensed subspace, and so on.  It's easy to
 prove~\cite{SM} that, if Eq.~(\ref{QPT0}) is satisfied, the $\TDlim$
 of $\min K_\mathrm{norm}/ \min K$ is 1, therefore, condition (i) is
 satisfied if in the $\TDlim$
 $\max V_\mathrm{cond}/N_{p} < \overline{V}/N_{p}$,
 i.e., $\max V_\mathrm{cond}\leq \overline{V}-\delta V$, with
 $\delta V>0$ being an arbitrary $O(N_{p})$ term.

 Condition (ii) is equivalent to saying that in the $\TDlim$
 $\min V_\mathrm{cond}/N_{p} = \min V/N_{p} < \min
 V_\mathrm{norm}/N_{p}$. Since in the $\TDlim$ we have
 $\min V_\mathrm{norm}/N_{p} = \max
 V_\mathrm{cond}/N_{p}$, the condition amounts to require
 $\max V_\mathrm{cond}/N_{p} > \min V/N_{p}$, i.e.,
 $\max V_\mathrm{cond} \geq \min V + \delta V$, $\delta V>0$ being an
 arbitrary $O(N_{p})$ term.

 In conclusion, the existence of any one of the partitions
 $\mathbb{F}=\mathbb{F}_\mathrm{cond} \oplus \mathbb{F}_\mathrm{norm}$
 obtained choosing
 $\min V + O(N_{p}) \leq \max V_\mathrm{cond} \leq \overline{V}
 - O(N_{p})$ allows us to say that, provided the screening
 length $R$ is finite, both Eqs.~(\ref{QPT0}) and (\ref{QCP}) are
 satisfied. It follows that the Hamiltonian $H=K+gV$ of
 Eqs.~(\ref{K}-\ref{g}) undergoes a Wigner crystallization in the form
 of a first-order QPT of the type introduced in~\cite{QPT}, i.e., as a
 condensation in the space of states.  About the critical parameter
 $g_{c}$, at this level we just know that it is finite.  The
 following of the Letter is devoted to the construction of upper and
 lower bounds of $g_{c}$ and, in order to do so, we shall
 exploit the invariance of the $\TDlim$~(\ref{gc_gcross}) under
 different partitions of $\mathbb{F}$.
 
 For finite $N$ and $N_{p}$, since $E_\mathrm{cond}$ and
 $E_\mathrm{norm}$ are monotonously increasing functions of $g$ convex
 upward, we have
 \begin{align}
   \label{gcross_bounds}
   g^-_\mathrm{cross} \leq g_\mathrm{cross} \leq g^+_\mathrm{cross},
 \end{align}
 where $g^+_\mathrm{cross}$ is the intersection point of two curves
 which are, respectively, a majorant of $E_\mathrm{cond}$ and a
 minorant of $E_\mathrm{norm}$, whereas $g^-_\mathrm{cross}$ is the
 intersection point of two curves which are, respectively, a minorant
 of $E_\mathrm{cond}$ and a majorant of $E_\mathrm{norm}$.  Indicating
 with $g^\pm_{c}$ the $\TDlim$s of $g^\pm_\mathrm{cross}$, we
 then have $g^-_{c} \leq g_{c} \leq g^+_{c}$. The
 more accurate are the approximations to $E_\mathrm{cond}$ and
 $E_\mathrm{norm}$, the tighter are the bounds
 $g^\pm_{c}$. However, we also want to choose these
 approximations to $E_\mathrm{cond}$ and $E_\mathrm{norm}$
 sufficiently simple to allow for an analytical evaluation of the
 $\TDlim$ of $g^\pm_\mathrm{cross}$.

 Let us examine the following inequalities
 \begin{align}
   \label{Econd+}
   E_\mathrm{cond} (g) & \leq g \min V_\mathrm{cond},
   \\
   \label{Enorm-}
   E_\mathrm{norm}(g) & \geq \min K_\mathrm{norm} + g \min
                        V_\mathrm{norm},
 \end{align}
 and
 \begin{align}
   \label{Econd-}
   E_\mathrm{cond}(g) & \geq \min K_\mathrm{cond} + g \min
                        V_\mathrm{cond},
   \\
   \label{Enorm+}
   E_\mathrm{norm}(g) & \leq \min K_\mathrm{norm} + g \max
                        V_\mathrm{norm}.
 \end{align}
 Equations~(\ref{Enorm-}), (\ref{Econd-}) and (\ref{Enorm+}) are
 Weyl's inequalities~\cite{MatrixTheory} for the lowest eigenvalue of
 $H=K+gV$ restricted to the condensed and normal subspaces.
 Equation~(\ref{Econd+}) follows from
 $E_\mathrm{cond} \leq \braket{u}{H}{u}/\scp{u}{u}$,
 $\forall\ket{u}\in\mathbb{F}_\mathrm{cond}$, choosing
 $\ket{u}=\ket{n}$, where $\ket{n}$ is any GS of $V$, and observing
 that $\braket{n}{K}{n}=0$.  From the first and second pair of
 inequalities we obtain, respectively,
 \begin{align}
   \label{gcross+}
   g^+_\mathrm{cross} = \frac{-\min K_\mathrm{norm}} {\min
   V_\mathrm{norm}-\min V_\mathrm{cond}},
 \end{align}
 \begin{align}
   \label{gcross-}
   g^-_\mathrm{cross} = \frac{\min K_\mathrm{cond}-\min
   K_\mathrm{norm}} {\max V_\mathrm{norm}-\min V_\mathrm{cond}}.
 \end{align}

 Consider Eq.~(\ref{gcross+}). We have $\min V_\mathrm{cond} = \min V$
 relying only on the filling $\varrho$ and the screening length $R$,
 the other quantities depend also on the choice of the condensed
 space.  We choose $\mathbb{F}_\mathrm{cond}$ in order to make
 $g^+_\mathrm{cross}$ as small as possible. A way is to make the
 denominator, therefore $\min V_\mathrm{norm}$, as large as possible.
 We assume
 $\min V_\mathrm{norm}/N_{p} = \max
 V_\mathrm{cond}/N_{p} \to \overline{V}/N_{p}$. In the
 numerator of~(\ref{gcross+}) we use
 $\min K_\mathrm{norm}/K_0\to 1$~\cite{SM}, where $K_0$ is the GS
 energy of $K$, namely,
 \begin{align}
   \label{K0}
   K_0 \equiv \min K = -2 \sin(\pi N_{p}/N) / \sin(\pi/N).
 \end{align}
 We thus obtain
 \begin{align}
   \label{gc+}
   g^+_{c} = \frac{-K_0/N_{p}}
   {\overline{V}/N_{p}-\min V/N_{p}}.
 \end{align}

 Consider Eq.~(\ref{gcross-}). We have already discussed
 $\min V_\mathrm{cond}$, as for $\max V_\mathrm{norm} = \max V$, it is
 the potential corresponding to the configurations in which the
 $N_{p}$ electrons are as tighter as possible, i.e., they
 occupy $N_{p}$ consecutive lattice sites. Thus the denominator
 of Eq.~(\ref{gcross-}) only depends on the filling $\varrho$ and the
 screening length $R$.  Now we choose $\mathbb{F}_\mathrm{cond}$ as
 small as possible, namely,
 $\max V_\mathrm{cond} \to \min V_\mathrm{cond}$. As before,
 $\min K_\mathrm{norm}/K_0 \to 1$.  We can also put
 $\min K_\mathrm{cond}/N_{p} \to 0$ as the number of allowed
 hoppings in $\mathbb{F}_\mathrm{cond}$ is, with this choice of
 $\max V_\mathrm{cond}$, at most $O(1)$.  Therefore
 \begin{align}
   \label{gc-}
   g^-_{c} = \frac{-K_0/N_{p}}{\max V/N_{p} - \min
   V/N_{p}}.
 \end{align}

 Equations (\ref{gc+}) and (\ref{gc-}) provide rigorous bounds to
 $g_{c}$.  From Table \ref{maindata}, it follows that, at
 filling $\varrho=3/10$ and screening length $R=10a$, a QPT takes
 place, in terms of the parameter $r_s$~\cite{rs}, at a critical value
 ${r_s}_{c}=g_{c}/4\varrho$ in the range
 $0.7 \leq {r_s}_{c} \leq 4.5$.
 \begin{table}
   \caption{\label{maindata} $\TDlim$ of the energies entering
     Eqs.~(\ref{gc+}) and (\ref{gc-}) and resulting bounds
     $g^\pm_{c}$ obtained at filling
     $\varrho=N_{p}/N=3/10$ and screening length
     $R=10a$~\cite{dimer}.}
   \begin{ruledtabular}
     \begin{tabular}{cccccc}
       ${\min V}/{N_{p}}$
       &
         ${\overline{V}}/{N_{p}}$
       &
         ${\max V}/{N_{p}}$
       &
         ${K_0}/{N_{p}}$
       &
         $g^+_{c}$
       &
         $g^-_{c}$
       \\
       $0.3846$
       &
         $0.7056$
       &
         $2.3518$
       &
         $-1.7168$
       &
         $5.4$
       &
         $0.84$
     \end{tabular}
   \end{ruledtabular}
 \end{table}

 In principle, $g_{c}$ could be estimated numerically by
 Eqs.~(\ref{QCPfinite}-\ref{gc_gcross}), allowing also for a direct
 evidence of the invariance of the choice of
 $\mathbb{F}_\mathrm{cond}$.  In fact, for different values of
 $\max V_\mathrm{cond}$ in the range allowed, we should observe
 different $g_\mathrm{cross}(N,N_{p})$ converging to the same
 $g_{c}$ in the $\TDlim$.  However, due to the growing speed of
 the Hilbert space, this program appears hopeless by standard
 numerical methods unless one uses ad hoc MC simulations.

\begin{figure}[t]
   \centering
   \includegraphics[width=1.0\columnwidth,clip]{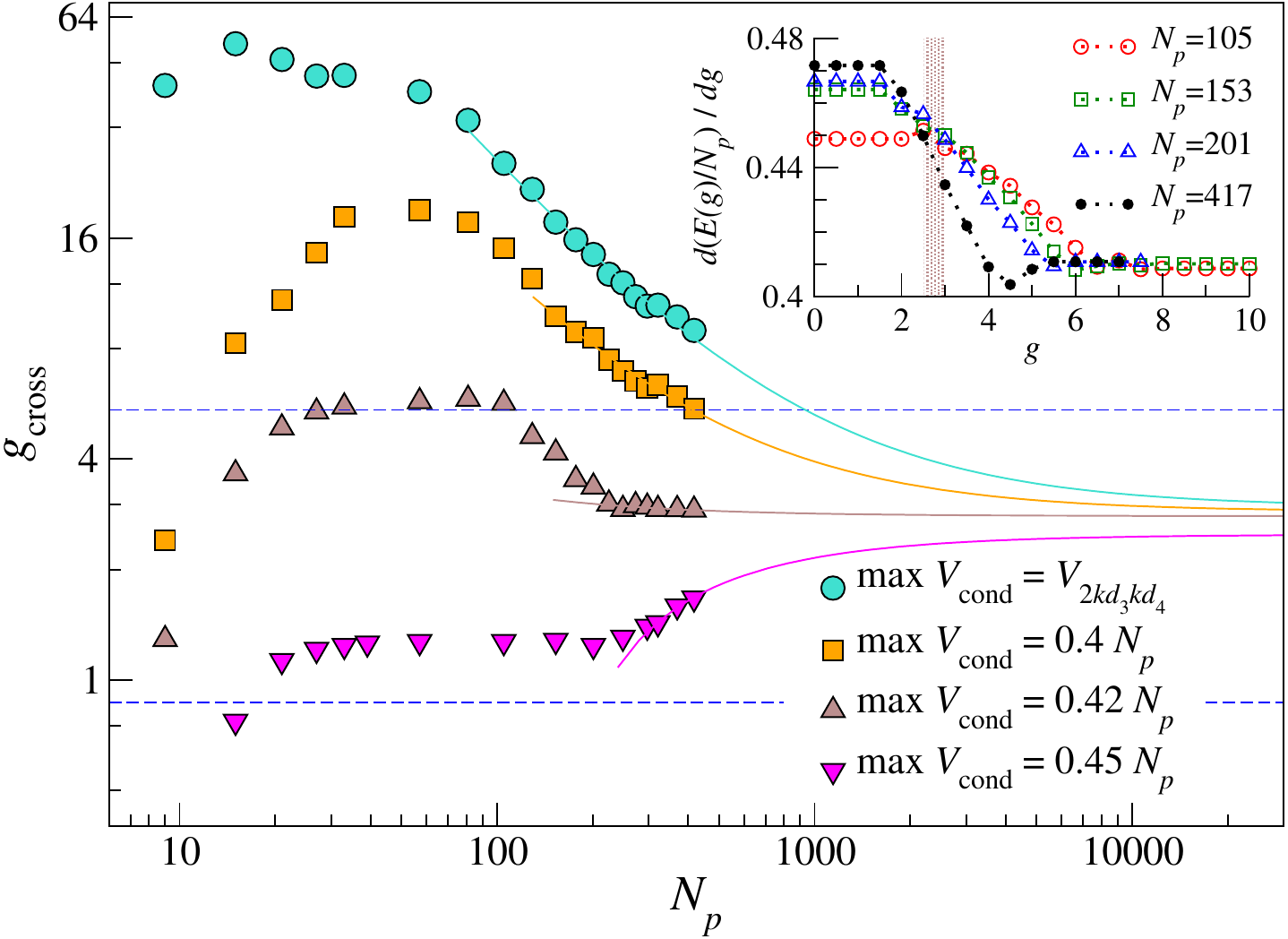}
   \caption{Value of $g_\mathrm{cross}(N,N_{p})$, solution of
     Eq.~(\ref{QCPfinite}), as a function of $N_{p}$ with
     filling $N_{p}/N=3/10$ and screening length $R=10 a$.
     Four different condensed subspaces (different
     $\max V_\mathrm{cond}$) are considered~\cite{V2kd3kd4}.
     Numerical MC data (symbols) are extrapolated to
     $N_{p}\to \infty$ by fitting $A+B/N_{p}$ (solid
     lines) to the points with largest $N_{p}$.  The horizontal
     dashed lines are the rigorous bounds of Eqs.~(\ref{gc+}) and
     (\ref{gc-}).  Inset: derivative of the GS energy per particle
     versus $g$ for different values of $N_{p}$. The shaded
     column indicates the value of $g_{c}=2.76 \pm 0.24$
     extrapolated as explained above.  }
   \label{fig_gcross_Np}
 \end{figure}

 We wrote a highly parallelized version, see \cite{SM} for details, of
 the bias-free MC algorithm derived from an exact probabilistic
 representation of the quantum evolution operator~\cite{EPR,MC}, and
 run it in a computer farm with thousands of nodes.  This allowed us
 to reach the remarkable size $N_{p}=417$, $N=1390$ with a
 computation time of several days per point, a point being the
 evaluation of $E_\mathrm{cond}(g)$ or $E_\mathrm{norm}(g)$ for a
 single value of $g$ and for a chosen system size.  The resulting
 values of $g_\mathrm{cross}(N,N_{p})$, at constant filling
 $\varrho=N_{p}/N=3/10$ and screening length $R=10a$, are shown
 in Fig.~\ref{fig_gcross_Np} as a function of $N_{p}$ for
 different choices of $\max V_\mathrm{cond}$.  Despite the very slow
 convergence of $g_\mathrm{cross}(N,N_{p})$ to $g_{c}$,
 note that the plot is shown in a log-log scale, all data sets appear
 to converge to a common $g_{c}$ whose value is within the
 rigorous bounds given before.  To estimate $g_{c}$, we fit the
 simple curve $A+B/N_{p}$ to the data obtained for large values
 of $N_{p}$, separately for each $\max V_\mathrm{cond}$.  The
 found values of $A$ suggest convergence to
 $g_{c}=2.76 \pm 0.24$ (i.e.,
 ${r_s}_{c} = 2.3 \pm 0.2$).  The first order nature of the QPT
 is made evident in the inset of Fig.~\ref{fig_gcross_Np}, where we
 report $d(E(N,N_{p},g)/N_{p})/dg$ versus $g$ for
 different values of $N_{p}$. By increasing $N_{p}$, we
 observe a developing discontinuity around the above estimate of
 $g_{c}$.  As a further signal of consistency, the derivatives
 of the GS energy tend to intersect toward a common point $g$ close to
 $g_{c}$~\cite{Binder}.

 Finally, we consider the limit $R\gg a$ in which screening becomes
 negligible.  In this limit we are able to express the characteristic
 potential values, {namely, $\min V/N_{p}$,
   $\max V/N_{p}$ and $\overline{V}/N_{p}$ in a closed
   analytical form~\cite{SM} We stress that these expressions are
   derived by first taking the $\TDlim$ and then picking the leading
   term for $R\gg a$.  By plugging these expressions together with
   $K_0/N_{p} \simeq -2\sin(\pi\varrho)/(\pi\varrho)$, obtained
   from Eq.~(\ref{K0}) for $R\gg a$, into Eqs.~(\ref{gc+}) and
   (\ref{gc-}), we find~\cite{rs}
   \begin{align}
     \label{rsc+-asymptotic}
     \frac{\sin(\pi\varrho)/(2\pi\varrho^2)} {\ln(R/a)-\varrho\ln(\varrho
     R/a)} \leq {r_s}_{c} \leq
     \frac{\sin(\pi\varrho)/(2\pi\varrho^2)}{-\varrho\ln(\varrho)} .
   \end{align}
   Equation~(\ref{rsc+-asymptotic}) allows us to estimate the
   dependence of ${r_s}_{c}$ on $\varrho$ in the range
   $a/R <\varrho \leq 1$, which, in virtue of the condition $R\gg a$,
   as a matter of fact coincides with the whole filling range.

   In the limit $R/a\to \infty$, the lower bound of
   Eq.~(\ref{rsc+-asymptotic}) vanishes whereas the upper bound
   remains finite. This is compatible with, but does not prove that
   ${r_s}_{c}\to 0$ in the limit of infinitely large screening
   length.  However, from Weyl's inequality
   $\min V (g+\min K/ \min V) \leq \min H \leq \min V (g+\max K/ \min
   V)$ and using the $R\gg a$ expressions of $K_0$ and $\min V$, we
   find
   \begin{align}
     \lim_{R/a\to\infty}
     \TDlim
     \frac{E(g)}{N_{p}}=\left\{
     \begin{array}{l}
       +\infty, \quad \quad \quad ~g>0, \\
       -2\frac{\sin(\pi\varrho)}{\pi\varrho}, \quad g=0.
     \end{array}
     \right.
   \end{align}
   We conclude that, if the $\TDlim$ is taken first, the Wigner
   crystallization is always realized as a first-order QPT of the
   type~\cite{QPT} but the critical parameter
   ${r_s}_{c}\to 0^+$ in the limit in which the potential
   becomes unscreened, $R/a\to\infty$.

\begin{acknowledgments}
  We are indebted to an anonymous referee of~\cite{QPT} for suggesting
  that we consider the Wigner crystallization (actually, as a
  counterexample of the present QPT mechanism).  Grant No. CNPq
  307622/2018-5 is acknowledged.  M.~O. thanks the Istituto Nazionale
  di Fisica Nucleare, Sezione di Roma 1, and the Department of Physics
  of Sapienza University of Rome for financial support and
  hospitality.  We thank Cineca, Consorzio Interuniversitario per il
  Calcolo Automatico, for access to its supercomputing facilities. We
  also thank Professor E. H. Lieb for letting us know of
  Ref.~\cite{Lieb2002}.
\end{acknowledgments}

% \bibliography{therm} % run pdflatex bibtex pdfltex pdflatex
% \input{this_file_name.bbl} % to be used before submission

%

\cleardoublepage
\newpage
\pagenumbering{arabic} \setcounter{page}{1} \onecolumngrid
\appendix

\begin{center}{\large\bf Supplemental Material for \\
    Wigner Crystallization of Electrons in a One-Dimensional Lattice:\\
    A Condensation in the Space of States}
\end{center}

\begin{center}
  Massimo Ostilli and Carlo Presilla
\end {center}

\setcounter{equation}{0}
\renewcommand{\theequation}{S\arabic{equation}}
\setcounter{figure}{0}
\renewcommand{\thefigure}{S\arabic{figure}}

\section{Ground states of $K$ and $V$}

For a ring of $N$ sites, the dimensionless hopping Hamiltonian is
\begin{align}
  K &= -\sum_{i=1}^{N}
      \left(c^\dag_{i}c_{i+1}+c^\dag_{i+1}c_{i}\right).
\end{align}
with $c_{i+N}=c_i$.  The GS of $K$, $\ket{K_0}$, is the product state
of the $N_{p}$ single-particle states with the lowest
single-particle energies among $\epsilon_n=-2\cos(2\pi n/N)$,
$n=0,\dots,N-1$.  For $N_{p}$ odd the corresponding GS energy
is
\begin{align}
  \label{SMK0}
  K_0 = \min K = -2 \sin(\pi N_{p}/N) / \sin(\pi/N).
\end{align}

In the same ring, the dimensionless interaction potential reads
\begin{align}
  \label{He}
  V = \sum_{i=1}^{N}\sum_{j=i+1}^{N} v_{i,j} c^\dag_i c_i c^\dag_j
  c_j=\sum_{i=1}^{N}\sum_{j=i+1}^{N} v_{i,j} n_i n_j,
\end{align}
with
\begin{align}
  v_{i,j} = \frac{1}{d_{i,j}} e^{-ad_{i,j}/R},
  \label{vij_SM}
\end{align}
where $d_{i,j}$ is the dimensionless distance between sites $i$ and
$j$ in the ring
\begin{align}
  \label{ring}
  d_{i,j} = \min(j-i,N+i-j), \quad j>i.
\end{align}
At filling $\varrho=p/q$, with $p$ and $q$ coprimes, there are $q$
degenerate classical WCs, i.e., $q$ configurations
$(n_1,n_2,\dots,n_N)$, with $n_i=0,1$ and
$\sum_{i=1}^{N}n_i=N_{p}$, which realize the minimum value of
the potential (\ref{He}).  For $p=1$ these are configurations with
equidistant fermions~\cite{SM_Fratini2004}, while for $p>1$ we have a
dimer structure~\cite{SM_Hubbard1978,Slavin2005}. For instance, at
filling $\varrho=3/10$, we have
$\min V=V_{k( d_3   d_3  \mathrm{d}_4)}$, which is the
potential of the 10 nonequivalent configurations obtained by repeating
$k=N_{p}/3=N/10$ times the sequence
$  d_3   d_3  \mathrm{d}_4$, where $ d_3 $ and
$ d_4 $ are the so called 3-dimers $(\circ,\circ,\bullet)$ and
4-dimers $(\circ,\circ,\circ,\bullet)$, namely, lattice segments of 3
or 4 sites in which only the last one is occupied.

\section{Characteristic values of the screened Coulomb energy}
The Wigner Crystallization cannot be clearly understood without an
analysis of the distribution of the values of the classical potential.
 An example of this distribution is given in
  Fig.~\ref{fig_potential_distribution}.
\begin{figure}[t]
  \centering
  \includegraphics[width=0.6\columnwidth,clip]{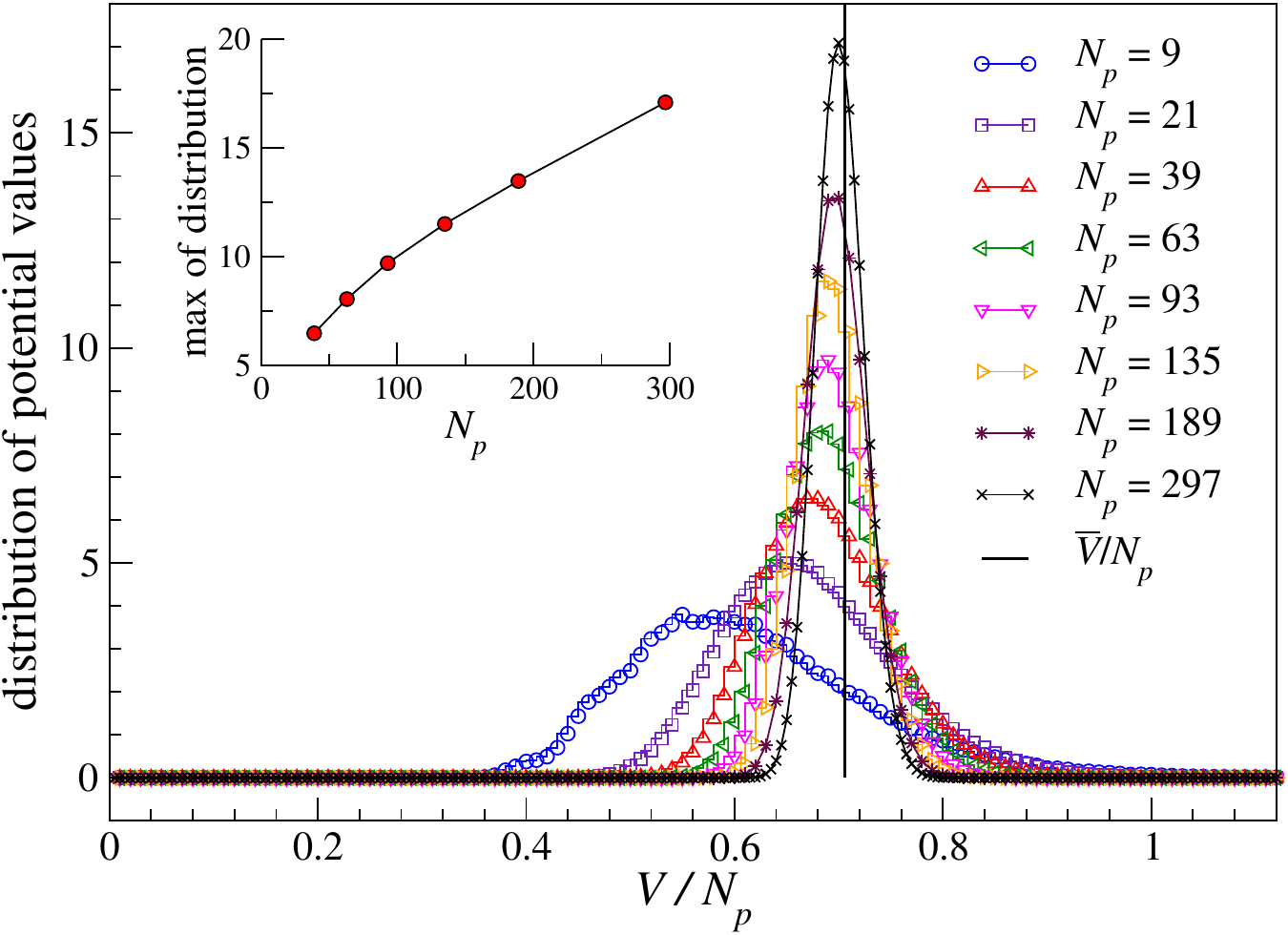}
  \caption
  {  Distribution of $V/N_{p}$ from Eq.~(\ref{vij_SM})
      with screening length $R=10 a$ and filling $\varrho=3/10$
      obtained by random sampling up to $2^{26}$ configurations for
      different values of $N_{p}$. The vertical line at
      $\overline{V}/N_{p}$ indicates the Dirac delta
      distribution obtained in the $\TDlim$.  Inset: maximum value of
      the distribution as a function of $N_{p}$.  }
  \label{fig_potential_distribution}
\end{figure}

In the following, we evaluate the $\TDlim$ necessary for the
implementation of the equations of the Letter:
$\min V/N_{p}$, $\max V/N_{p}$, the classical mean value
${\overline{V}}/{N_{p}}$, and the gap (see Eqs. (31)-(34) of
the Letter).  As we shall see, compact formulas can be provided in
the important limit of large screening length, $R\gg a$.  We shall
also demonstrate why the distribution of $V/{N_{p}}$ tends to a
Dirac delta distribution centered on the mean value.

\subsection{Minimum of $V$}
Let us evaluate the minima $\min V$, i.e., the value of $V$ evaluated
in any of its $q$ GSs (we recall that the filling is $\varrho=p/q$
with $p$ and $q$ coprimes).  The GS of $V$ takes the following
expression
\begin{align}
  \label{MinV}
  & \min V = \sum_{j=1}^{N_{p}-1} v_{1,r(j)}
    +\sum_{j=r(1)}^{N_{p}-1} v_{r(1),r(j)}+\ldots, ~\\
  \label{rj}
  & r(j)=1+[N/N_{p}]j+[(N/N_{p}-[N/N_{p}])j],
\end{align}
where $[\cdot]$ stands for integer part and $r(j)$ represents the
position of the $(j+1)$th particle~\cite{SM_Slavin2005}.  When
$N/N_{p}$ is an integer (i.e., when $p=1$), we have
$r(j)=1+N/N_{p} j$ and in the ring, where the distance is
defined as in (\ref{ring}), the above expression simplifies neatly in
\begin{align}
  \label{MinV0}
  & \min V = \frac{N_{p}}{2} \sum_{j=1}^{N_{p}-1}
    v_{1,r(j)}.
\end{align}
When instead $N/N_{p}$ is not an integer, the approximation
$r(j)\simeq 1+N/N_{p} j$ will produce in general not integer
values above and below the corresponding exact values of $r(j)$. For
large values of $N$ and $N_{p}$ with $N_{p}/N$ fixed,
these above and below approximations will tend to be somehow
distributed around the exact values of $r(j)$. In such a limit we can
hence still use Eq. (\ref{MinV0}) as an approximation and, from the
explicit expression of the $v_{i,j}$ we get (we suppose $N_{p}$
odd as in the Letter)
\begin{align}
  \label{MinV1}
  \min V \simeq N_{p} \sum_{j=1}^{(N_{p}-1)/2}
  \frac{e^{-d_{1,r(j)}a/R}}{d_{1,r(j)}},
\end{align}
Note that in the above summation we have only $d_{i,j}=j-i$.  In the
following, we shall apply the same approximation to all the other
terms, i.e., we will take $r(j)\simeq 1+jN/N_{p}$ (which
amounts to $d_{1,r(j)}\simeq jN/N_{p}$) thoroughly. By using
the variable $x=Naj/N_{p}R$, we get (now we take
$N_{p}-1\simeq N_{p}$)
\begin{align}
  \label{MinV2}
  \min V \simeq N_{p} \frac{a}{R}
  \sum_{x=Na/N_{p}R}^{Na/2R} \frac{e^{-x}}{x},
\end{align}
and
\begin{align}
  \label{MinV3}
  \TDlim \frac{\min
  V}{N_{p}} \simeq \frac{a}{R} \sum_{x=a/\varrho
  R}^{\infty} \frac{e^{-x}}{x},
\end{align}
which in turn gives
\begin{align}
  \label{MinV4}
  \lim_{R/a \gg 1}
  \TDlim \frac{\min
  V}{N_{p}} \simeq \varrho \int_{a/\varrho
  R}^{\infty} dx \frac{e^{-x}}{x}.
\end{align}
The above integral can be split as a part over the interval
$[1,\infty]$ and a part over the interval $[a/\varrho R,1]$. The
former is a finite dimensionless constant $I_1$, whereas the latter
gives
\begin{align}
  \label{MinV5}
  \int_{a/\varrho R}^{1} dx \frac{e^{-x}}{x}= \ln(\varrho R/a) + I_2.
\end{align}
where $I_2$ is another finite constant. In conclusion, we have
\begin{align}
  \label{MinV6}
  \lim_{R/a \gg 1}
  \TDlim \frac{\min
  V}{N_{p}} \simeq \varrho \ln(\varrho R/a)+\mathop{O}(1).
\end{align}
In general, in the limit $R/a\to\infty$ the term $\mathop{O}(1)$ is a
small correction that depends on $\varrho$ and is exactly zero for
densities such that $N/N_{p}$ is an integer. See
Fig.~\ref{fig_Vasymptotic}.

\subsection{Maximum of $V$}
Let us calculate $\max V$, i.e., the maximum value of $V$ evaluated in
any of the $N$ ways that exist to put the $N_{p}$ particles in
$N_{p}$ consecutive sites. We have
\begin{align}
  \label{MaxV}
  \max V = v_{1,2} + & v_{1,3}+v_{1,4}+\ldots + v_{1,N_{p}}+ ~ 
                       \nonumber\\
                     & v_{2,3} + v_{2,4} + \ldots + v_{2,N_{p}}+
                       ~\nonumber\\
                     & \quad \quad \quad \quad ~ + \ldots + ~\nonumber\\
                     & \quad \quad \quad \quad \quad \quad ~~ +
                       v_{N_{p}-1,N_{p}},
\end{align}
which, by using $v_{1,2}=v_{2,3}$, etc., gives
\begin{align}
  \label{MaxV1}
  \max V = \sum_{n=1}^{N_{p}-1}(N_{p}-n)v_{1,n+1}.
\end{align}
We can now proceed as in the previous case arriving at
\begin{align}
  \label{MaxV2}
  \lim_{R/a \gg 1}
  \TDlim \frac{\max
  V}{N_{p}} \simeq
  \left[\ln(R/a)+\mathop{O}(1)\right] -\lim_{R/a \gg 1}
  \TDlim
  R \frac{e^{-a/R}}{N_{p}},
\end{align}
or
\begin{align}
  \label{MaxV3}
  \lim_{R/a \gg 1}
  \TDlim \frac{\max
  V}{N_{p}} \simeq \ln(R/a)+\mathop{O}(1) .
\end{align}
Also in this case, in the limit $R/a\to\infty$ the term
$\mathop{O}(1)$ is a small correction that depends on $\varrho$ and is
exactly zero for densities such that $N/N_{p}$ is an
integer. See Fig.~\ref{fig_Vasymptotic}.

\subsection{Classical mean value and distribution of the normalized
  values $V/N_{p}$}
The classical mean value of $V$ is defined as the average over all
configurations.  If we indicate by $\overline{~\cdot~}$ these
averages, from Eq. (\ref{He}) we have
\begin{align}
  \label{MeanV0}
  \overline{V}=\frac{\sum_{n}\langle
  n|V|n\rangle}{M}=\sum_{i=1}^{N}\sum_{j=i+1}^{N} v_{i,j}
  \overline{n_i n_j}.
\end{align}

In Eq. (\ref{MeanV0}) one is tempted to neglect correlations and to
replace $\overline{n_i n_j}$ with
$\overline{n_i} \cdot\overline{n_j}=\varrho^2$. In this way we get
\begin{align}
  \label{MeanV01}
  \frac{\overline{V}}{N_{p}}\simeq
  \frac{\varrho^2}{N_{p}}\sum_{i=1}^{N}\sum_{j=i+1}^{N}
  v_{i,j}.
\end{align}
It turns out that this approximation becomes exact in the
thermodynamic limit, however, the reason for that is quite not
trivial.  By analyzing the distribution of the classical
$V/N_{p}$ values in the thermodynamic limit, we can
simultaneously understand why Eq. (\ref{MeanV01}) becomes exact and
why the distribution tends to a Dirac delta distribution centered at
${\overline{V}}/{N_{p}}$.

In general, $\min V$ is $q$-fold degenerate, whereas $\max V$ is
$N$-fold degenerate, and as we consider more and more intermediate
values of $V$, the degeneracy grows exponentially fast with the system
size.  This can be better understood in terms of entropy. Let us first
consider the case $\varrho=1/q$ and let us split the $N$ sites into
$N/q=N_{p}$ segments each made up of $q$ sites. Let us
enumerate these segments from $j=1$ to $N_{p}$.  In each
segment, we can accommodate a number of particles $m_j$ between 0 and
$q$ provided that the constrain $\sum_j m_j=N_{p}$ is
satisfied.  There are many possible ways to realize a given sequence
of segments $\{m_j\}$ and the corresponding potential $V$ might be
different for each one of such realizations.  We are interested in
counting the total number of configurations $\mathcal{N}(\{m_j\})$
associated to a given sequence of segments $\{m_j\}$, independently of
the different values of $V$.  Taking into account that the particles
are indistinguishable and double occupancy of a site is forbidden, the
number of configurations $\mathcal{N}(\{m_j\})$ associated to a given
sequence of segments $\{m_j\}$ is
\begin{align}
  \label{MeanV}
  \mathcal{N}\Large(\{m_j\}\Large)=\delta\left(\sum_j
  m_j-N_{p}\right)\prod_j \binom{q}{m_j},
\end{align}
For $N$ large, by a small variation of the sequence of segments we can
have a large variation of $\mathcal{N}\large(\{m_j\}\large)$.  In
fact, not surprisingly, it is easy to check that,
$\ln(\mathcal{N}\large(\{m_j\}\large))$, i.e., the canonical entropy,
is exponentially peaked around its maximum which is attained by the
uniform segment distribution, $\{m_j=1\}$.  The important point here
is that, at $\{m_j=1\}$, $V$ still depends on the particular
realization of the uniform segment distribution and we are precisely
interested in evaluating the mean value of these values of $V$ because
in correspondence of the uniform segment distribution $\{m_j=1\}$
there are concentrated the most frequent values of $V$ (in fact, as
the entropy shows, exponentially more frequent than the other values).
For these values we have $V\in [\min V,\max V|_{\{m_j=1\}}]$, where
$\min V$, as we already know, is obtained by putting, for example, all
the particles in the rightmost position of the segments, and
$\max V|_{\{m_j=1\}}$ is obtained by putting, for example, the
particle of the $i$-th segment on the leftmost position of the segment
when $i$ is odd, and on the rightmost position when $i$ is
even. Notice that, for $\varrho<1/2$ (as thoroughly supposed in our
work,) $\max V|_{\{m_j=1\}}$ is strictly lower than $\max V$ (it is
easy in particular to evaluate it in the limit of large screening
length, where we get
$\max V|_{\{m_j=1\}}/N_{p}\simeq 2\varrho \ln(R/a)$, to be
compared with $\max V/N_{p}\simeq \ln(R/a)$ from
Eq.~(\ref{MaxV3})).  Notice also that $\min V$, which is $q$-fold
degenerate, and $\max V|_{\{m_j=1\}}$, which is $2q$-fold degenerate,
are just two extremal values of the uniform segment distribution
$\{m_j=1\}$ but they are not typical. The typical values of $V$ in the
range $[\min V,\max V|_{\{m_j=1\}}]$ are more complicated and must
have exponential degeneracies (or, more in general, nearly
degeneracies).

We have so far shown that, in the $\TDlim$, the distribution of
$V/N_{p}$ tends to a Dirac delta distribution centered on its
mean value which in turn must coincide with the mean value restricted
to the uniform segment distribution, $\{m_j=1\}$.  It is this latter
fact that allows us to evaluate the mean of $V/N_{p}$ in a
simple way: since in each segment we have exactly one particle, we are
no more concerned with correlations so that the replacement
$\overline{n_i n_j}$ with
$\overline{n_i} \cdot\overline{n_j}=\varrho^2$, as in
Eq. (\ref{MeanV01}), is actually exact in the $\TDlim$. In other
words, we can approximate the $N_{p}$ segments as uniformly
occupied by a continuous distribution of charge of density
$\varrho=1/q$.  In particular, in the limit of large screening length
Eq. (\ref{MeanV01}) provides
\begin{align}
  \label{MeanV1}
  \lim_{R/a \gg 1}
  \TDlim
  \frac{\overline{V}}{N_{p}} \simeq \varrho \ln(R/a)+\mathrm{O}(1).
\end{align}
The above result has been derived for simplicity in the case
$\varrho=1/q$, however, the same arguments can be equally repeated in
the general case of $\varrho=p/q$ and the result is still
Eqs. (\ref{MeanV01}) and (\ref{MeanV1}). See
Fig.~\ref{fig_Vasymptotic}.

\begin{figure}[h]
  \centering
  \includegraphics[width=0.6\columnwidth,clip]{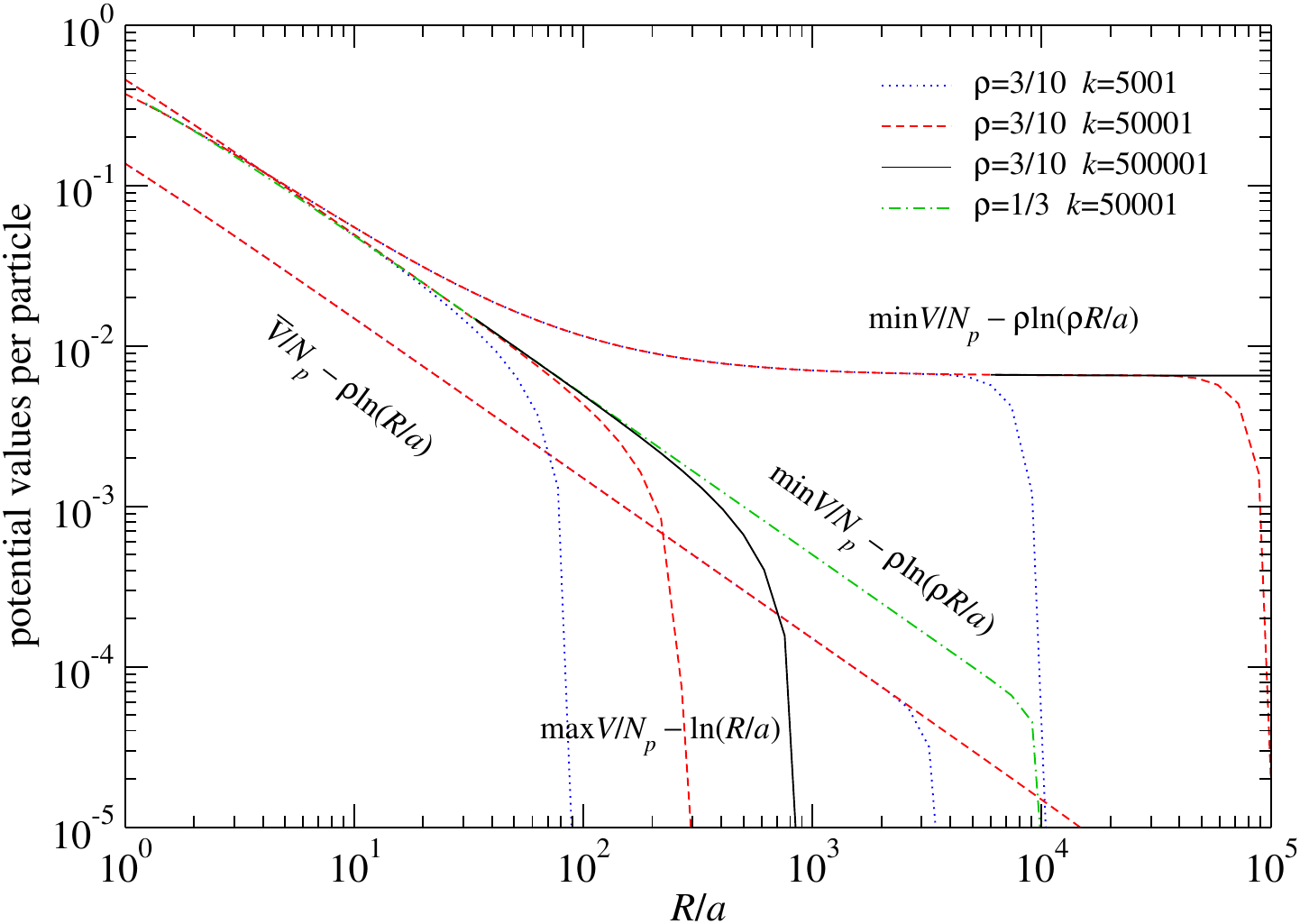}
  \caption{The exact minimum, maximum, and classical mean value of the
    potential $V$ per particle compared with the leading terms of
    Eqs. (\ref{MinV6}), (\ref{MaxV3}), and (\ref{MeanV1}),
    respectively, as a function of the screening length $R/a$ for
    filling $\varrho=3/10$ and $\varrho=1/3$.  The integer
    $k=1,2,\ldots$ determines the size of the system: we have
    $N_{p}=3k$ and $N=10k$ for filling $3/10$ and
    $N_{p}=k$ and $N=3k$ for filling $1/3$.  Provided the
    system size is sufficiently large ($\TDlim$ almost reached), when
    $1/\varrho$ is an integer, the difference between the exact values
    and the leading terms is always $O(1/(R/a))$ (here we show only
    the curve relative to $\min V$).  When $1/\varrho$ is not an
    integer, the curve corresponding to $\min V$ tends to a small
    $\mathrm{O}(1)$ term.  All curves show a drop for $R/a$
    sufficiently large, meaning that the $\TDlim$ cannot be considered
    reached at the used $k$ values for the shown values of $R/a$.}
  \label{fig_Vasymptotic}
\end{figure}

\section{Gap of $V$}
The gap is defined as the difference between the first excited value
of $V$ and its minimum. If $1/\varrho$ is integer (i.e., $p=1$), the former
can be obtained by shifting one single particle in the GS of $V$ by
one hop toward a first neighbor vacant position. By using the
definitions (\ref{ring}) and (\ref{rj}), we have
\begin{align}
%  \label{alpha}
  \mathrm{gap}(V) =
  \sum_{j=1}^{N_{p}~-1}v_{2,r(j)}-\sum_{j=1}^{N_{p}-1}v_{1,r(j)}=
  \sum_{j=1}^{(N_{p}-1)/2}v_{2,r(j)}+
  \sum_{j=(N_{p}+1)/2}^{N_{p}-1}
  v_{2,r(j)}-2\sum_{j=1}^{(N_{p}-1)/2}v_{1,r(j)}\nonumber.
\end{align}
On defining the variables $x=(jN/N_{p})a/R$,
$x_1=(jN/N_{p}-1)a/R$, and $x_2=[N-(jN/N_{p}-1)]a/R$, we
obtain
\begin{align}
  \label{alpha1}
  \mathrm{gap}(V) =
  \frac{a}{R}\left[\sum_{x_1=\min x_1}^{\max x_1}
    \frac{e^{-x_1}}{x_1}+\sum_{x_2=\min x_2}^{\max x_2}\frac{e^{-x_2}}{x_2}
    -2\sum_{x=\min x}^{\max x}\frac{e^{-x}}{x}\right],
\end{align}
where
\begin{align*}
%  \label{alpha2}
  &\min x=\frac{1}{\varrho}\frac{a}{R}, \qquad 
  \max x=\frac{N_{p}-1}{2\varrho}\frac{a}{R}, \\
  &\min x_1=\frac{1-\varrho}{\varrho}\frac{a}{R}, \qquad 
  \max x_1=\frac{N_{p}-1-2\varrho}{2\varrho}\frac{a}{R}, \\
  &\min x_2=\frac{2}{\varrho}\frac{a}{R}, \qquad 
  \max x_2=\frac{N_{p}+1}{2\varrho}\frac{a}{R}.
\end{align*}
In the limit of large screening length we get
\begin{align}
  \label{alpha3}
  \lim_{R/a \gg 1} \TDlim
  \mathrm{gap}(V)\simeq
  \varrho\int_{\frac{1-\varrho}{\varrho}
      \frac{a}{R}}^{\infty}dx_1\frac{e^{-x_1}}{x_1}+
    \varrho\int_{\frac{2}{\varrho}\frac{a}{R}}^{\infty}dx_2\frac{e^{-x_2}}{x_2}
  -2
  \varrho\int_{\frac{1}{\varrho}\frac{a}{R}}^{\infty}dx\frac{e^{-x}}{x} ,
\end{align}
which gives
\begin{align}
  \label{alpha4}
  \lim_{R/a \gg 1} \TDlim
  \mathrm{gap}(V)\simeq -\varrho\ln[2(1-\varrho)].
\end{align}
Whereas for the cases $1/\varrho$ integer the above formula turns out
to be almost exact (e.g., for $\varrho=1/3$, Eq. (\ref{alpha4}) gives
$\varrho\ln[2(1-\varrho)]=0.0958$, while the exact value of the gap is
$0.0986$), when $1/\varrho$ is not an integer, it provides only a rough
approximation since, for such cases, the first excited state of
$V$ cannot be obtained by simply shifting one single particle in the
GS of $V$.  In particular, for $\varrho=3/10$ Eq. (\ref{alpha4})
gives a value about 50 times larger than the actual value.  In
general, the first excited state of $V$ corresponds to a not trivial
modification of its GS.

\section{Monte Carlo simulations}

For finite size systems we can evaluate several properties of the GS
by means of Monte Carlo simulations, other numerical methods being
excluded due to the huge size of the Hilbert space.  In the following,
we discuss data obtained by an unbiased Monte Carlo code~\cite{SM_MC1}
based on an exact probabilistic representation of the quantum
evolution operator~\cite{SM_EPR1}. Note that we always simulate
systems with an odd number of fermions in order to avoid any sign
problem~\cite{SM_Wiese}.  The code has been parallelized using openMP.

The relevant code parameters~\cite{SM_MC1} that we used in our
simulations are: $2^{14}$ stochastic trajectories (independent Poisson
processes), $2^{12}$ reconfigurations and a time $3/N_{p}$
between consecutive reconfigurations (corresponding to about 10 jumps
of the Poisson processes).  For the largest simulated system with 417
particles in a lattice of 450 sites, this required a computation time
of about 200 hours per single $g$ point for each subspace. Since the
crossing between $E_\mathrm{cond}(g)$ and $E_\mathrm{norm}(g)$ can be
obtained by simulating both these two GS energies in, at least, 2 $g$
points, we obtain a computation time for $g_\mathrm{cross}$ of about
800 hours.  The absence of bias effects is checked by evaluating $E$
at $g=0$ and comparing the result with $K_0$, the GS energy of $K$,
for which we have the explicit formula (\ref{SMK0}). The total
computation time at this size is, in conclusion, about 1000 hours.

In Fig.~\ref{fig_energies_vs_g} we show the behavior of the energies
per particle, $E_\mathrm{cond}/N_{p}$,
$E_\mathrm{norm}/N_{p}$ and $E/N_{p}$, as a function of
$g$ in the case $N_{p}=15$, $N=50$ and with the choice
$\max V_\mathrm{cond} = 0.4 N_{p}$. It is evident that $E$
interpolates between $E_\mathrm{norm}$ at $g$ small and
$E_\mathrm{cond}$ at $g$ large. The functions $E_\mathrm{cond}(g)$ and
$E_\mathrm{norm}(g)$ intersect at $g_\mathrm{cross} \simeq 8.3$.  Note
that, whereas the values of $E_\mathrm{cond}$, $E_\mathrm{norm}$ and
$g_\mathrm{cross}$ depend on the choice of the condensed subspace,
univocal thermodynamic limits $\eps_\mathrm{cond}$,
$\eps_\mathrm{norm}$ and $g_{c}$ are obtained for any allowed
$\mathbb{F}_\mathrm{cond}$.

Figure 2 of the Letter is obtained by using also a suitable importance
sampling which turns out to be effective at large values of $g$.
Concerning the Inset of Fig. 2, where we evaluate
$d(E(N,N_{p},g)/N_{p})/dg$, we have made use of the
Savitsky-Golay filter~\cite{SM_SG} applied to the MC data in order to
smooth the otherwise too noisy signal (by first applying the filter to
the MC data and then evaluating the derivative, or by directly
applying the filter to evaluate the derivative produce similar
results; the Inset shows the latter).

\begin{figure}
  \centering
  \includegraphics[width=0.6\columnwidth,clip]{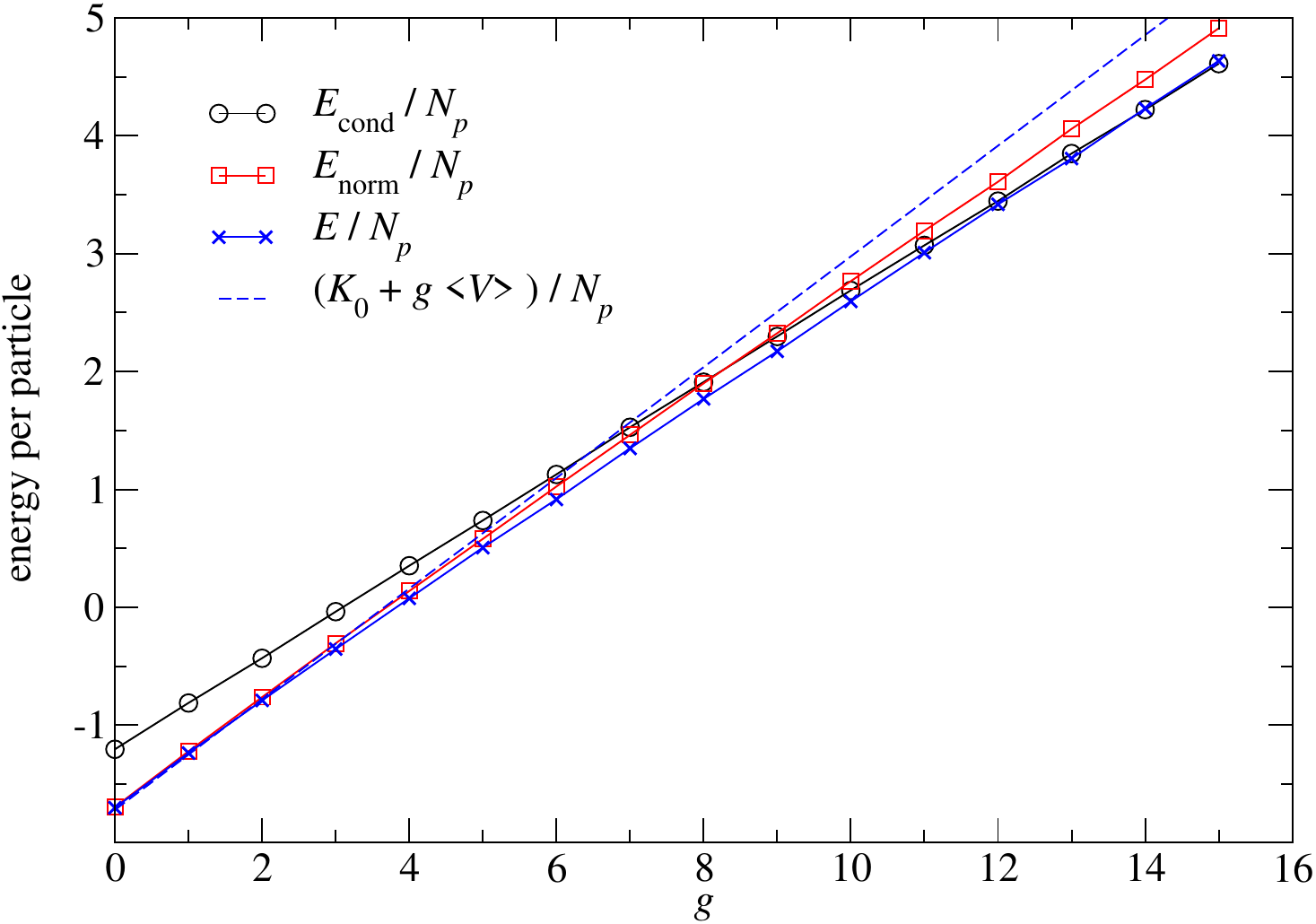}
  \caption{Energies per particle, $E_\mathrm{cond}/N_{p}$,
    $E_\mathrm{norm}/N_{p}$ and $E/N_{p}$, for the
    Hamiltonian $H=K+gV$ as a function of $g$.  We have
    $N_{p}=15$, $N=50$ and screening length $R=10 a$.  As
    condensed subspace we use that defined by
    $V_\mathrm{cond} = 0.4 N_{p}$. The dashed line is the GS
    energy of $H$ obtained, at the first order of perturbation theory,
    considering $gV$ as a perturbation of $K$.}
  \label{fig_energies_vs_g}
\end{figure}

\section{Perturbation theory}
We have not made use of any finite perturbation theory.  The following
represents only a complementary study that could be used for
consistency.

For $g$ small, we can approximate the energy $E$ of the GS of $H$ by
using the first order perturbation theory. We have
$E = K_0 + g \media{V}$, where $\media{V} = \braket{K_0}{V}{K_0}$ is
\begin{align}
  \label{mediaV}
  \media{V} &= \sum_{i=1}^{N} \sum_{j=i+1}^{N}
              \frac{e^{-ad_{i,j}/R}}{d_{i,j}}
              \left(\frac{N_{p}}{N}\right)^2 \nonumber \\ 
              &\qquad\times
              \left[ 1- \left( \frac{\sin(\pi N_{p}(j-i)/N)}
                  {N_{p}\sin(\pi(j-i)/N)} \right)^2\right].
\end{align}
In terms of limiting rescaled energies we thus conclude that for $g$
small (see Fig. \ref{fig_energies_vs_g})
\begin{align*}
  \epsilon_\mathrm{norm} \simeq
  \TDlim
  \frac{K_0}{N_{p}} + g
  \TDlim
  \frac{\media{V}}{N_{p}}.
\end{align*}

\textit{Remark}.  The GS of $K+gV$ tends to $\ket{K_0}$, or to one of
the classical WCs, in the limits $g\to 0 $ and $g\to\infty$,
respectively. Now, whereas for $g$ sufficiently small is safe to
assume that the actual GS is a slight deformation of $\ket{K_0}$,
i.e., a product state of single particle Bloch waves, the
investigation of the actual GS for $g$ sufficiently large is quite
more complex. In fact, depending on $p$, different ansatzs have been
proposed in the past: a Bloch superposition of kink-antikink
configurations for $p=1$~\cite{SM_Fratini2004}, and of excited dimers for
$p>1$~\cite{SM_Slavin2005}.  However, whereas these ansatzs provide
physical appealing insights, they remain heuristic as essentially
focus on single mode excitations. There is no reason to exclude
\textit{a priori} that an extensive number of kink-antikink walls
($p=1$), or other nondimer configurations ($p>1$) concur to the
actual GS for $g$ finite.  In fact, in both cases, the potential
values associated to the configurations contributing to these Bloch
states differ from the energy of the WCs (i.e., the minima of $V$) by
terms $O(1)$, while our condition (ii) on $\mathbb{F}_\mathrm{cond}$
requires a larger space, being
$\max V_\mathrm{cond}=\min V + O(N_{p})$, and only under such a
condition a QPT can be reached.

\section{Monotonicity and convexity of the GS energies}
Consider the Taylor expansion and the infinite perturbation series of
the GS energy $E(g)$ both around an arbitrary $g$ and compare term by
term the first- and second-order terms of the two expansions.  By
using the fact that the first-order term of the perturbation series is
$\langle E(g)| V |E(g)\rangle/\langle E(g)|E(g)\rangle \geq 0$, we get
$\partial E(g)/\partial g \geq 0$ (this result can be equally reached
by using the Hellman-Feynman theorem).  Next, by using the fact that
the second-order term of the perturbation series for the GS energy is
always negative or null, we also get $\partial^2 E(g)/\partial g^2\leq
0$. The same argument applies to $E_\mathrm{cond}$ and
$E_\mathrm{norm}$. In conclusion, with respect to $g$, all the GS
energies are functions that are monotone increasing and convex upward.

\section{Proof that Eq. (9) implies $\min K_\mathrm{norm}/K_0= 1$ in
  the $\TDlim$}
The starting point is the exact probabilistic representation of the
quantum evolution introduced in~\cite{SM_EPR1}. According to this exact
representation, at an imaginary time $t$, we have
\begin{align}
  \label{EPR0}
  \braket{\bm{n}}{e^{-Ht}}{\bm{n}_{0}} =  
  \E  \left(  
  \mathcal{M}^{[0,t)}_{\bm{n}_{0}} \delta_{ \bm{n}_{N_t} , \bm{n} }
  \right), 
\end{align}
where $\E(\cdot)$ is the probabilistic expectation over the continuous
time Markov chain of configurations
$\bm{n}_{0},\bm{n}_{s_1},\dots,\bm{n}_{s_{N_t}}$ (or trajectory)
defined by the transition matrix
\begin{align}
  \label{Pf} 
  P_{\bm{n},\bm{n}'} = \frac{|\braket{\bm{n}}{K}{\bm{n}'}|}{A(\bm{n})},
  \qquad A(\bm{n}) = \sum_{\bm{n}'} |\braket{\bm{n}}{K}{\bm{n}'}|,
\end{align}
and the sequence of jumping times $s_1,s_2,\dots,s_{N_t}$ obtained
from the Poissonian conditional probability density
\begin{align}
  P(s_{k}|s_{k-1}) =
  e^{-\Gamma A(\bm{n}_{s_{k-1}}) (s_{k}-s_{k-1})}
  \Gamma A(\bm{n}_{s_{k-1}}).
  \label{Ps}
\end{align}
Concretely, starting form the configuration $\bm{n}_{0}$ at time
$s_0=0$, we draw a configuration $\bm{n}_{s_1}$ with probability
$P_{\bm{n}_{0},\bm{n}_{s_1}}$ at time $s_1$ drawn with probability
density $P(s_{1}|s_{0})$, then we draw a configuration $\bm{n}_{s_2}$
with probability $P_{\bm{n}_{s_1},\bm{n}_{s_2}}$ at time $s_2$ drawn
with probability density $P(s_{2}|s_{1})$, and so on until we reach
the configuration $\bm{n}_{N_t}$ at time $s_{N_t}$ such that
$s_{N_t+1}>t$. Note that the Poisson processes associated to each jump
are defined left continuous~\cite{SM_EPR1}, as a consequence, the
configuration $\bm{n}_{N_t}$ is the one realized by the Markov chain
just before the final time $t$.  The stochastic functional
$\mathcal{M}^{[0,t)}_{\bm{n}_0}$ is then defined as
\begin{align}
  \label{EPR}
  \mathcal{M}^{[0,t)}_{\bm{n}_0}
  &= e^{\sum_{k=0}^{N_t-1} [\Gamma A(\bm{n}_{s_k})
    -V(\bm{n}_{s_k})](s_{k+1}-s_{k})}\cdots
    e^{[\Gamma A(\bm{n}_{s_{N_t}})
    -V(\bm{n}_{s_{N_t}})](t-s_{N_t})},
\end{align}
where $A(\bm{n})$ is called the number of links, or degree of
$\bm{n}$, and represents the number of non-null off-diagonal matrix
elements $\langle \bm{n} |H|\bm{n}' \rangle$.

The exact probabilistic representation (\ref{EPR0}) is at the base of
the unbiased Monte Carlo simulations used to sample the GS properties
(by sending the imaginary time $t$ to sufficiently large values), as
explained before (see Figs. S2 and S3). Eq. (\ref{EPR0}), however,
lends itself also to quite interesting analytical treatments allowing
for a direct connection between GS properties and the (virtual)
trajectories of the Markov chain~\cite{SM_ANAL,ANAL1}.  Here we consider
the case with no interaction $V\equiv 0$ and $N_{p}$ odd, i.e.,
a lattice chain of spinless fermions with no sign problem (equivalent
to a system of hard-core bosons).  As done in \cite{SM_ANAL}, in this
particular case we can easily decompose the expectation of the
stochastic functional $\mathcal{M}^{[0,t)}_{\bm{n}_0}$ as a sum over
trajectories that, starting from a given initial configuration
${\bm{n}_{0}}$, perform $N_t=k$ jumps within the time $t$. By
integrating out the ordered jumping times $s_1,s_2,\dots,s_{k}$
distributed according to the probability density (\ref{Ps}), and by
using
$\int_{0}^t \Gamma ds_1\int_{s_1}^t \Gamma ds_2\ldots \int_{s_{k-1}}^t
\Gamma ds_{k}=(\Gamma t)^{k}/(k!)$, we arrive at
\begin{align}
  \label{EPR1}
  \sum_{\bm{n}}\braket{\bm{n}}{e^{-Ht}}{\bm{n}_{0}}=
  \E  \left( \mathcal{M}^{[0,t)}_{\bm{n}_{0}}\right)=
  \sum_k^{\infty} \frac{(\Gamma t)^{k}}{k!}
  \mathcal{N}(\bm{n}_{0};k),
\end{align}
where $\mathcal{N}(\bm{n}_{0};k)$ counts the total number of
trajectories having $k$ jumps (each starting from $\bm{n}_{0}$).
Similarly, for the Hamiltonians $H_{\mathrm{cond}}$ and
$H_{\mathrm{norm}}$ defined as the Hamiltonian $H$ restricted to the
subspaces $\mathbb{F}_{\mathrm{cond}}$ and
$\mathbb{F}_{\mathrm{norm}}$, respectively, we have
\begin{align}
  \label{EPRcond}
  \sum_{\bm{n}\in \mathbb{F}_{\mathrm{cond}}}
  \braket{\bm{n}}{e^{-H_{\mathrm{cond}}t}}{\bm{n}_{\mathrm{cond}}}
  =\sum_k^{\infty} \frac{(\Gamma t)^{k}}{k!}
  \mathcal{N}_{\mathrm{cond}}(\bm{n}_{\mathrm{cond}};k),
\end{align}
\begin{align}
  \label{EPRnorm}
  \sum_{\bm{n}\in \mathbb{F}_{\mathrm{norm}}}\braket{\bm{n}}
  {e^{-H_{\mathrm{norm}}t}}{\bm{n}_{\mathrm{norm}}}=
  \sum_k^{\infty} \frac{(\Gamma t)^{k}}{k!}
  \mathcal{N}_{\mathrm{norm}}(\bm{n}_{\mathrm{norm}};k),
\end{align}
where $\bm{n}_{\mathrm{cond}}$ and $\bm{n}_{\mathrm{norm}}$ are two
arbitrary initial configurations of $\mathbb{F}_{\mathrm{cond}}$ and
$\mathbb{F}_{\mathrm{norm}}$, respectively.  On expanding the lhs of
these equations to leading order in $t$ we get
\begin{align}
  \label{EPR1b}
  \sum_{\bm{n}}\langle \bm{n}|\bm{n}_{0}\rangle e^{-Et}=
  \sum_k^{\infty} \frac{(\Gamma t)^{k}}{k!}
  \mathcal{N}(\bm{n}_{0};k),
\end{align}
\begin{align}
  \label{EPRcondb}
  \sum_{\bm{n}\in \mathbb{F}_{\mathrm{cond}}}\langle\bm{n}|
  \bm{n}_{\mathrm{cond}}\rangle e^{-E_{\mathrm{cond}}t}
  =\sum_k^{\infty} \frac{(\Gamma t)^{k}}{k!}
  \mathcal{N}_{\mathrm{cond}}(\bm{n}_{\mathrm{cond}};k),
\end{align}
\begin{align}
  \label{EPRnormb}
  \sum_{\bm{n}\in \mathbb{F}_{\mathrm{norm}}}\langle\bm{n}|
  \bm{n}_{\mathrm{norm}}\rangle e^{-E_{\mathrm{norm}}t}=
  \sum_k^{\infty} \frac{(\Gamma t)^{k}}{k!}
  \mathcal{N}_{\mathrm{norm}}(\bm{n}_{\mathrm{norm}};k).
\end{align}
On the other hand, taking into account that
$\mathcal{N}_{\mathrm{cond}}(\bm{n}_{\mathrm{cond}};k)$ and
$\mathcal{N}_{\mathrm{norm}}(\bm{n}_{\mathrm{norm}};k)$ are fast
growing functions of the space dimensions $M_{\mathrm{cond}}$ and
$M-M_{\mathrm{cond}}$, respectively, if condition (\ref{QPT0}) is
satisfied, $\TDlim {M_\mathrm{cond}/M}=0$, independently from the
choice of the initial configurations, we clearly see that, for system
size sufficiently large, and for any $k$ larger than some finite
threshold
\begin{align}
  \label{simple}
  \mathcal{N}_{\mathrm{norm}}(\bm{n}_{\mathrm{norm}};k)>
  \mathcal{N}_{\mathrm{cond}}(\bm{n}_{\mathrm{cond}};k).
\end{align}
Comparing Eqs. (\ref{EPRcondb}) and (\ref{EPRnormb}) we therefore
conclude that, for system size sufficiently large,
\begin{align}
  \label{simpleb}
  E_{\mathrm{norm}}<E_{\mathrm{cond}},
\end{align}
which, plugged into Eq. (\ref{QPT1}), proves that
$\min K_\mathrm{norm}/K_0= 1$ in the $\TDlim$.


\begin{thebibliography}{99}%

\bibitem{Wigner} E.~Wigner, On the interaction of electrons in metals,
  Phys. Rev. \textbf{46}, 1002 (1934).

\bibitem{Hubbard1978} J.~Hubbard, Generalized Wigner lattices in one
  dimension and some applications to tetracyanoquinodimethane(TCNQ)
  salts, Phys. Rev. B \textbf{17}, 494 (1978).

\bibitem{Sinai1983} S.~E.~Burkov and Y.~G.~Sinai, Phase diagrams of
  one-dimensional lattice models with long-range antiferromagnetic
  interaction, Russ. Math. Surv. \textbf{38}, 235 (1983).

\bibitem{Bak1982} P.~Bak and R.~Bruinsma, One-Dimensional Ising Model
  and the Complete Devil's Staircase, Phys. Rev. Lett. \textbf{49},
  249 (1982).

\bibitem{Fratini2004} S.~Fratini, B.~Valenzuela, and D.~Baeriswyl,
  Incipient quantum melting of the one-dimensional Wigner lattice,
  Synth. Met. \textbf{141}, 193 (2004).

\bibitem{Slavin2005} V.~Slavin, Low-energy spectrum of one-dimensional
  generalized Wigner lattice, Phys. Status Solidi B \textbf{242}, 2033
  (2005).


\bibitem{2DEG_1989} B.~Tanatar and D.~M.~Ceperley, Ground state of the
  two-dimensional electron gas, Phys. Rev. B \textbf{39}, 5005 (1989).

\bibitem{2DEG_1996} F.~Rapisarda and G.~Senatore, Diffusion Monte
  Carlo study of electrons in two-dimensional layers,
  Aust. J. Phys. \textbf{49}, 161 (1996).

\bibitem{2DEG_2002} C.~Attaccalite, S.~Moroni, P.~Gori-Giorgi, and
  G.~B.~Bachelet, Correlation Energy and Spin Polarization in the 2D
  Electron Gas, Phys. Rev. Lett. \textbf{88}, 256601 (2002).

\bibitem{2DEG_2009} N.~D.~Drummond and R.~J.~Needs, Phase Diagram of
  the Low-Density Two-Dimensional Homogeneous Electron Gas,
  Phys. Rev. Lett. \textbf{102}, 126402 (2009).

\bibitem{2DEG_2017} M.~Zarenia, D.~Neilson, B.~Partoens, and
  F.~M.~Peeters, Wigner crystallization in transition metal
  dichalcogenides: A new approach to correlation energy, Phys. Rev. B
  \textbf{95}, 115438 (2017).

\bibitem{Noda2002} Y.~Noda and M.~Imada, Quantum Phase Transitions to
  Charge-Ordered and Wigner-Crystal States under the Interplay of
  Lattice Commensurability and Long-Range Coulomb Interactions,
  Phys. Rev. Lett. \textbf{89}, 176803 (2002).


\bibitem{Siegmund2009} M.~Siegmund, M.~Hofmann, and O.~Pankratov,
  Density functional study of collective electron localization:
  Detection by persistent current, J. Phys. Condens. Matter
  \textbf{21} 155602 (2009).

\bibitem{VFB2003} B.~Valenzuela, S.~Fratini, and D.~Baeriswyl, Charge
  and spin order in one-dimensional electron systems with long-range
  Coulomb interactions, Phys. Rev. B \textbf{68}, 045112 (2003).


\bibitem{nanotube} I.~Shapir, A.~Hamo, S.~Pecker, C.~P.~Moca,
  \"O~Legeza, G.~Zarand and S.~Ilani,Imaging the electronic Wigner
  crystal in one-dimension, Science, \textbf{364}, 870 (2019).


\bibitem{Dubin} D.~H.~E.~Dubin, Minimum energy state of the
  one-dimensional Coulomb chain, Phys. Rev. E \textbf{55}, 4017
  (1997).

\bibitem{Lieb2002} H.~J.~Brascamp, E.~H.~Lieb, Some inequalities for
  Gaussian measures and the long-range order of the one-dimensional
  plasma, edited by M.~Loss and M.~B.~Ruskai, in \textit{Inequalities}
  (Springer, Berlin, Heidelberg, 2002).

\bibitem{QPT} M.~Ostilli and C.~Presilla, First-order quantum phase
  transitions as condensations in the space of states, J. Phys. A
  \textbf{54}, 055005 (2021).
  
\bibitem{MC} M.~Ostilli and C.~Presilla, Exact Monte Carlo time
  dynamics in many-body lattice quantum systems, J. Phys. A
  \textbf{38}, 405 (2005).

\bibitem{rs} The relation between our parameter $g$ and the parameter
  $r_s$ usually appearing in the literature is as follows.  In a
  one-dimensional lattice of spacing $a$ with $N$ sites and
  $N_{p}$ electrons, the radius of the volume available to each
  electron is $r=(a N/N_{p})/2$. Therefore
  $r_s \equiv r/a_{B}= g/4\varrho$.

\bibitem{SM} See Supplemental Material at <url> for details, which
  includes Refs.~\cite{Wiese,SG,ANAL,ANAL1}.

\bibitem{Wiese} M. Troyer and U.-J. Wiese, Computational Complexity
  and Fundamental Limitations to Fermionic Quantum Monte Carlo
  Simulations, Phys. Rev. Lett. \textbf{94}, 170201 (2005).


\bibitem{SG} A. Savitzky and M. J. E. Golay, Smoothing and
  differentiation of data by simplified least squares procedures,
  Anal. Chem., \textbf{36}, 1627 (1964).
  
\bibitem{ANAL} M. Ostilli and C. Presilla, An analytical probabilistic
  approach to the ground state of lattice quantum systems: Exact
  results in terms of a cumulant expansion, J.  Stat. Mech., (2005)
  P04007.

\bibitem{ANAL1} M. Ostilli and C. Presilla, The exact ground state for
  a class of matrix Hamiltonian models: Quantum phase transition and
  universality in the thermodynamic limit, J. Stat. Mech., (2006)
  P11012.


  
\bibitem{MatrixTheory} J.~N.~Franklin \textit{Matrix Theory}, (Dover
  Publications, New York, 1993).

\bibitem{dimer} At this filling
  $\min V = V_{k( d_3   d_3  \mathrm{d}_4)}$, which is
  the potential associated to the so-called dimer configuration
  $ d_3   d_3  \mathrm{d}_4$ repeated
  $k=N_{p}/3=N/10$ times, see
  Refs. \cite{Hubbard1978,Slavin2005} and comments in Ref. \cite{SM}.
  
\bibitem{EPR} M. Beccaria, C. Presilla, G. F. De Angelis, and G. Jona
  Lasinio, An exact representation of the fermion dynamics in terms of
  Poisson processes and its connection with Monte Carlo algorithms,
  Europhys. Lett. \textbf{48}, 243 (1999).
  
\bibitem{Binder} K.~Binder, Finite Size Scaling Analysis of Ising
  Model Block Distribution Functions, Z. Phys. B \textbf{43}, 119-140
  (1981).

\bibitem{V2kd3kd4} The potential
  $\max V_\mathrm{cond} = V_{2k  d_3  k \mathrm{d}_4}$ is the
  potential of the most excited configuration with $2k$ dimers
  $ d_3 $ and $k$ dimers $\mathrm{d}_4$, where
  $k=N_{p}/3=N/10$.  The value
  $V_{2k  d_3  k \mathrm{d}_4}/N_{p}$ depends on
  $N_{p}$ but in $\TDlim$
  $V_{2k  d_3  k \mathrm{d}_4}/N_{p} \to 0.3925$.

  
\end{thebibliography}

\begin{thebibliography}{3}%
  
\bibitem{SM_Fratini2004} S.~Fratini, B.~Valenzuela, and D.~Baeriswyl,
  ``Incipient quantum melting of the one-dimensional Wigner lattice'',
  Synthetic Metals \textbf{141}, 193 (2004).

\bibitem{SM_Hubbard1978} J.~Hubbard, ``Generalized Wigner lattices in one
  dimension and some applications to tetracyanoquinodimethane(TCNQ)
  salts'', Phys. Rev. B \textbf{17}, 494 (1978).

\bibitem{SM_Slavin2005} V.~Slavin, ``Low-energy spectrum of
  one-dimensional generalized Wigner lattice'', Phys. Stat. Sol. (b)
  \textbf{242}, 2033 (2005).

\bibitem{SM_MC1} M.~Ostilli and C.~Presilla, ``Exact Monte Carlo time
  dynamics in many-body lattice quantum systems'', J. Phys. A
  \textbf{38}, 405 (2005).

\bibitem{SM_EPR1} M. Beccaria, C. Presilla, G. F. De Angelis, G. Jona
  Lasinio, ``An exact representation of the fermion dynamics in terms
  of Poisson processes and its connection with Monte Carlo algorithms
  '', Europhys. Lett. \textbf{48}, 243 (1999).

\bibitem{SM_Wiese} M. Troyer, U.-J. Wiese, ``Computational Complexity and
  Fundamental Limitations to Fermionic Quantum Monte Carlo
  Simulations'', Phys. Rev. Lett. \textbf{94}, 170201 (2005).


\bibitem{SM_SG} A. Savitzky, M. J. E. Golay, ``Smoothing and
  Differentiation of Data by Simplified Least Squares Procedures'',
  Analytical Chemistry, \textbf{36}, 1627 (1964).
  
\bibitem{SM_ANAL} M. Ostilli and C. Presilla, ``An analytical
  probabilistic approach to the ground state of lattice quantum
  systems: exact results in terms of a cumulant expansion'', J.
  Stat. Mech., P04007 (2005).

\bibitem{SM_ANAL1} M. Ostilli and C. Presilla, ``The Exact ground state
  for a class of matrix Hamiltonian models: quantum phase transition
  and universality in the thermodynamic limit'', J. Stat. Mech.,
  P11012 (2006).
  % DOI: 10.1088/1742-5468/2006/11/P11012
  
\end{thebibliography}
\end{document}